\documentclass[a4paper,twoside]{article}
\newcommand{\affil}[1]{$^{\rm #1}$}
\usepackage[authoryear]{natbib}
\bibpunct{(}{)}{;}{a}{}{,}
\usepackage{graphicx}
\date{} 
\newcommand{\ha}{\mbox{H$\alpha$}}
\newcommand{\hi}{\mbox{H{\sc I}}}
\newcommand {\hr}{\mbox{$^{h}$}}
\newcommand {\dd}{\mbox{$^{\circ}$}}
\newcommand {\m}{\mbox{$^{m}$}}
\newcommand {\s}{\mbox{$^{s}$}}
\title{\large\bf\flushleft \ha\ emission from the Magellanic Bridge}
\author{\parbox{\textwidth}{\flushleft
\vspace{-0.5cm}
{\it Muller, E.\affil{A}, Parker, Q.A.\affil{B,C}\\
\vspace{0.4cm}
{\small \affil{A}\,Bolton Fellow, Australia Telescope National Facility, CSIRO, P.O. Box 76, Epping, NSW 1710, Australia}\\
{\small \affil{B}\,Department of Physics, Macquarie University, Sydney, NSW 2109, Australia}\\
{\small \affil{C}\,Anglo-Australian Observatory, Epping, NSW 1710, Australia}\\
{\small Email: emuller@csiro.au}
}}}
\begin{document}
\twocolumn[
\begin{changemargin}{.8cm}{.5cm}
\begin{minipage}{.9\textwidth}
\vspace{-1cm}
\maketitle
%
%
\small{\bf Abstract:} We present here a preliminary report and commentary of recently processed observations of \ha\ emission towards the Magellanic Bridge. These data have been analysed in an attempt to quantify the extent to which the stellar population is capable of reshaping the local ISM. We find that the \ha\ emission regions are small, weak and sparsely distributed, consistent with a relatively quiescent and inactive ISM where radiative and collisional ionisation is inefficient and sporadic. This suggests that energetic processes at the small scale (i.e. $\sim$tens of pc) do not dominate the energy balance within the ISM of the Bridge, which therefore hosts a pristine turbulent structure, otherwise inaccessible within our own Galaxy. We find \ha\ emission that is well correlated with detected $^{12}$CO(1-0) line emission (a proxy for molecular hydrogen), as well as other easily identified ring-like \hi\ features. 


\medskip
\medskip
\end{minipage}
\end{changemargin}
]
\section{Introduction}
\begin{figure*}
\begin{center}
\includegraphics[scale=.5]{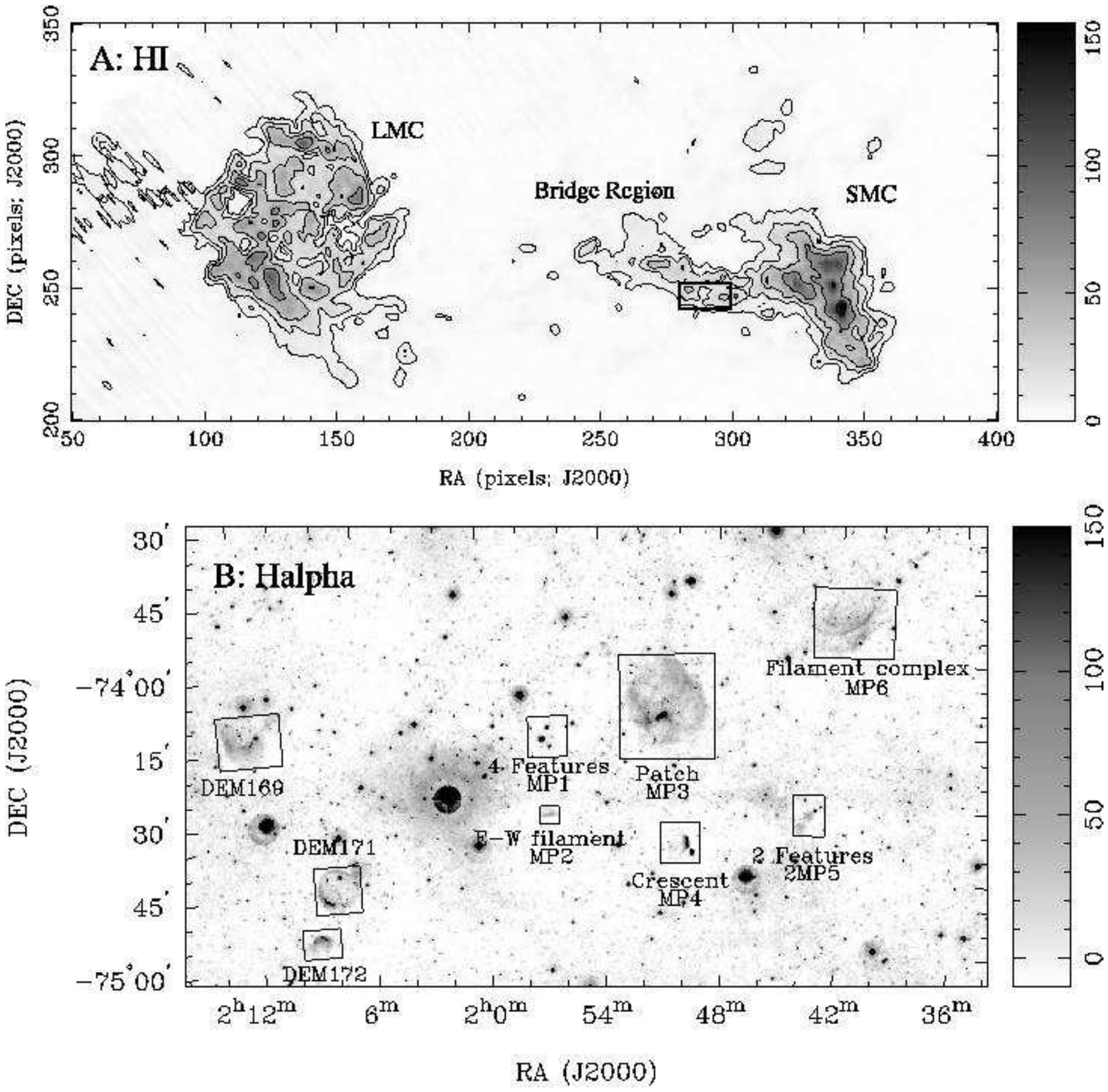}
\end{center}
  \caption{(\emph{Panel A}) Integrated \hi\ intensity of the Magellanic System \citep[data from][]{putman}. Contours are 5,10,20,40,80\% of peak intensity where the units are in Kelvin. The overlaid box indicates the extent of the selected \ha\ field used in this study, shown in Panel B.\newline
(\emph{Panel B}) A sub-region of the complete UKST \ha\ dataset is magnified and located against the \hi\ brightness distribution. NE is in the direction of the top left. Nine boxes have been positioned over identified \ha\ emission. The three easternmost boxes, labelled as DEMxxx, are known features from \cite{dem} \citep[see also ][]{meaburn}. The other features are newly discovered in this study. The intensity units are arbitrary, with a linear transfer function. See Figures \ref{fig:dem171}-\ref{fig:f_filaments} for more accurate flux estimates. The bright object at 2$^h$ 03$^m$ is a foreground star.}\label{fig:magsys}
\end{figure*}

The Magellanic Bridge comprises a streamer of \hi\ which has been gravitationally drawn from the Large and Small Magellanic Clouds (LMC, SMC) approximately 200 Myr ago, during a close  (3 kpc) grazing pass between the two Clouds \citep[e.g.][]{gardiner}. A sub-region of recent \hi\ measurements of the Magellanic System are shown in Figure~\ref{fig:magsys} \citep{putman}. The dense regions on the east (\emph{left}) and west (\emph{right}) sides of the Figure are the ISM of the Small and Large Magellanic Clouds and the Magellanic Bridge can be seen as an eastward extension of the SMC, stretching towards the LMC.

The ISM of the Magellanic System and the Bridge in particular, have come to be regarded as templates for early-epoch systems, due to their relatively un-enriched ISM and slow star formation rate. These properties, along with their relatively close proximity \citep[LMC: 50 kpc, SMC: 65 kpc, e.g.][]{abraha} and large subtended spatial length, make them ideal for high-resolution studies of the turbulent characteristics of a poorly enriched ISM.

A turbulent fluid flow behaves according to both a hierarchical scale-power distribution, and scale-dependent energy injection mechanisms \citep[e.g.][]{kolmogorov}. The canonical size-power relation for an unperturbed fluid conforms to a logarithmic law: P(k)$\propto$k$^{\gamma}$ where k is the wave number. For an incompressible and turbulent fluid flow, $\gamma$ is theoretically predicted to be $\sim$-3.6 \citep[e.g.][]{kolmogorov}. In general, broad-scale observations of the Galactic and Magellanic gaseous ISM yield values of $\gamma$ between $\sim$-2.2 to -3 \citep{green,elmegreen,mul_stats}.

\cite{maclow} has shown that the dominant energy injection mechanism into the ISM is via supernovae (SNe), although this is surely true only for systems which harbour a substantial, massive stellar population. The Magellanic Bridge is known to host a young stellar population \citep[e.g.][and references therein.]{bica,DI}, some of which are early types. The Bridge is also known to be an active, but inefficient producer of stars \citep{christo,mizuno}. As such, the ISM in the Bridge may constitute a textbook turbulent structure: one that is not significantly perturbed by SNe or stellar winds and which conforms well to the expected hierarchical spatial-power cascade \citep[e.g.][]{kolmogorov,deshpande}. 

The predictable turbulent structure of the Bridge ISM is certainly true for scales 10$^3$-10$^2$ pc according to current measurements of the pervasive neutral ISM  within the Magellanic Bridge \citep[i.e. \hi;][]{mul_hi}. However these observations are limited to the spatial resolutions accessible by the ATCA and it is currently impossible to sample the ISM down to scales $<$ $\sim$100 pc using those data. Due to the prohibitively large observation integration times, it is difficult to obtain higher resolution data at usable sensitivities, with the view to measuring the small-scale arrangement of the ISM. To estimate the energy balance at smaller (i.e. stellar) scales we therefore turn to higher resolution measurements of observable signatures of energetic processes.

We examine here the \ha\ emission from the Magellanic Bridge as a signpost of significant and localised energetic processes, in an attempt to estimate the energy balance at these relatively small spatial scales. 
Previous studies of the \ha\ emission from the Bridge is typically wide-field, but of poor resolution \citep{jmo}, or single-pointing, very targeted and of narrow field of view \citep{mbg,putman_ha}. To date, only one extended object has been subject to any dedicated morphological study at all: a well-formed ring of \ha\ emission that may be generated by a SN, or from a central 08-type star \citep[see Figure~\ref{fig:dem171}; e.g.][]{meaburn}. 

We have newly identified a small number of extended \ha\ features within the Magellanic Bridge. Many of these are comfortably attributable to radiative ionisation from bright UV sources and there is at least one example of a shock-ionised \ha\ feature.

\section{Data processing}
We made use of the existing archived data to obtain our \ha\ maps of the Magellanic bridge region. They were taken by the Anglo-Australian Observatory UK Schmidt Telescope, at Siding Spring, Australia as part of the Southern Galactic Plane and Magellanic Clouds (MC) \ha\ survey \citep{parker_ha}. Forty fields on overlapping 4~degree centres in and around the Magellanic Clouds were observed using a 305mm clear aperture \ha\ interference filter of exceptional quality \citep{parker_filter} and high resolution, high sensitivity Tech-Pan film as detector \citep{parker_film}. Each field was exposed for 3~hours yielding a nominal $\sim$5 Rayleigh sensitivity \citep{parker_ha}.
The original Magellanic Cloud survey fields were scanned by the SuperCOSMOS measuring machine \citep{hambly} and the 10$\mu$m pixel data have a resolution limit at 0.67$"$ (though of course typical seeing is 2$"$ and the smallest point source images are $\sim20-25\mu$m. Each \ha\ MC survey field subtends $\sim$5.5 degrees on the sky.

The target region for the data used here is shown as a rectangle overlaid on the wing feature in Figure~\ref{fig:magsys}, although the actual scope of the Schmidt field is much larger. The digitised SuperCOSMOS data have been flat fielded, calibrated and continuum subtracted according to \cite{parker_ha}. The lack of extended emission in the field, used to calibrate the emission for example via direct reference to SHASSA images (Gaustad et al. 2001), makes accurate calibration very difficult and generates uncertainties in flux of at least $\sim$20-30\%. All \ha\ data shown here (excluding Figure~\ref{fig:magsys}A) have been filtered with a small (3$\times$3 pixel) median filter to remove crowding and to highlight extended structure. Due to the variation of filter artifacts and the very low level of the \ha\ emission, the \ha\ data shown in Figure~\ref{fig:magsys}A have not been smoothed or filtered and the large-scale variation has been removed only by a low-order fit. As such, the intensity units of Figure~\ref{fig:magsys}A are unreliable and are given only as an approximately relative measure. The difficulties in obtaining accurate calibration and the large resulting potential errors have encouraged us to initiate future more detailed analysis of some of the more enigmatic and intriguing objects discussed in this initial report.

\section{The \ha\ features:}
A wide-field \ha\ image of the part of the Bridge containing the brightest \hi\ emission is shown in Figure~\ref{fig:magsys}A. Although the processing techniques will have removed any kpc-scale gradient across the field, extended \ha\ emission is detected in $\sim9$ smaller regions.

\begin{figure}
\begin{center}
  \includegraphics[scale=.3, angle=-90]{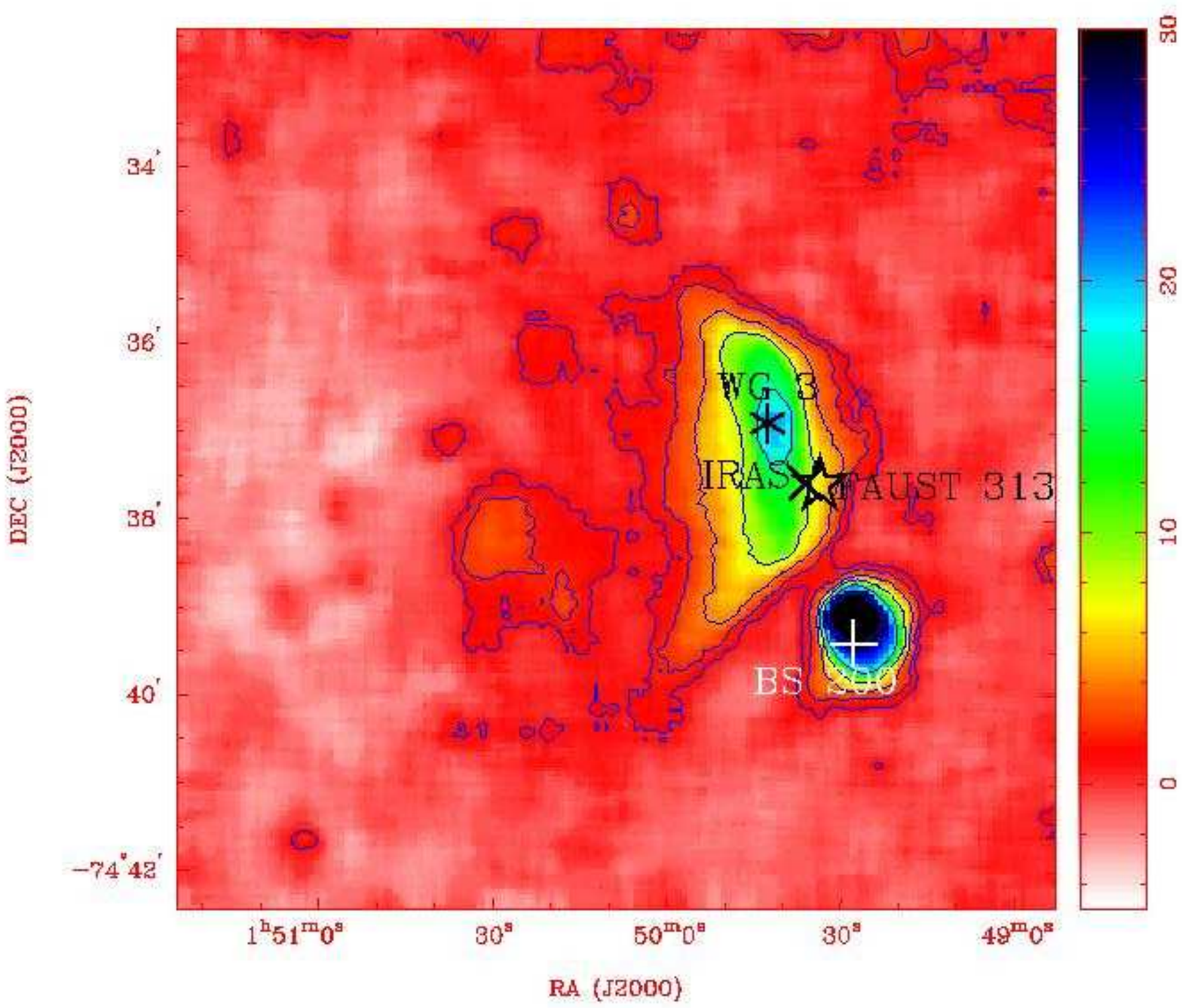}
\end{center}
  \caption{UKST H-alpha data, showing a crescent (MP4). This crescent-like feature in the Bridge is well associated with potential heating sources. FAUST 313 is shown with the Star symbol and the cluster WG 2 \citep{wg71} is shown with the black asterisk. The OB association BS95 200 is located with the white plus sign. Contours are 2$^x$\% of peak intensity ($x$=0-6). Intensity units shown on the colour wedge are in Rayleighs.}\label{fig:crescent}
\end{figure}

We show a magnification of each region in Figures \ref{fig:dem171}$-$\ref{fig:f_filaments}.  In general, the extended \ha\ is very weak, indicating that the ISM harbours a low mass of neutral hydrogen or that weakly ionising processes are in operation. Table 1 summarises the defining characteristics of each of the features and provides some speculation on the mechanisms responsible for ionising the ISM in the Bridge.

In all cases, we have attempted to correlate the projected positions of any putative sources of ionizing flux. FAUST objects \citep{faust}, OB stars and associations \citep{bd,bica,DI} that are found within 10'-20' (170-350 pc) of each \ha\ region are plotted on top of \ha\ emission.

Only two of the detected \ha\ features appear to exhibit a regular ring morphology (see Figures~\ref{fig:dem171},\ref{fig:crescent}). Other features are more amorphous, showing clumpy, filamentary or otherwise unstructured distributions of the \ha\ emission region. We will discuss each of the features below in some detail.

\subsection{Ring feature -DEM171}
\begin{figure}
\begin{center}
  \includegraphics[scale=.3, angle=-90]{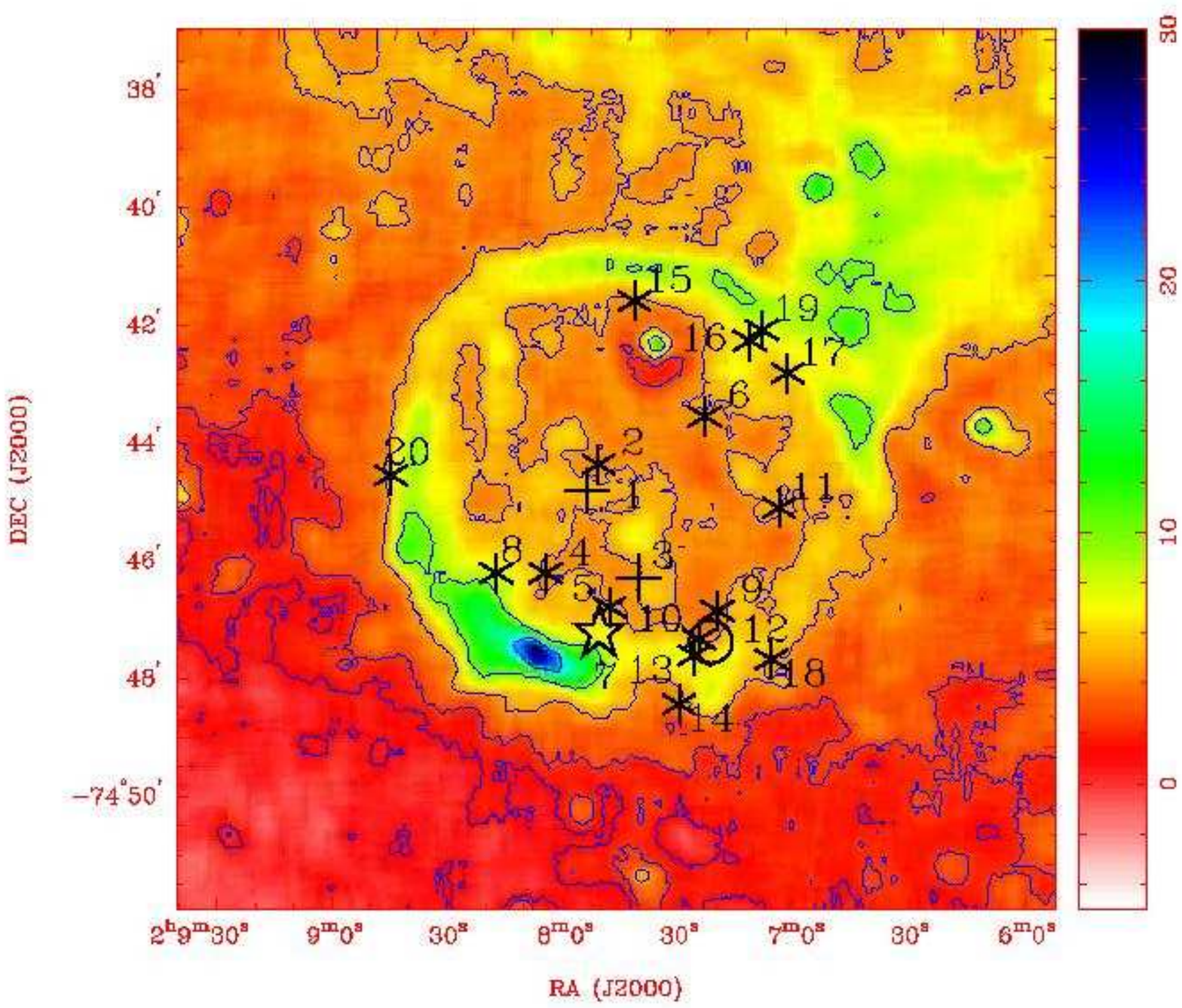}
\end{center}
  \caption{The \ha\ feature DEM171 observed from the relevant UKST H-alpha MC survey field that covers that part of the Bridge. This \ha\ shell is also coincident with an easily identifiable expanding HI shell \citep{mul_hi}. The overlaid symbols indicate different catalogue sources: an asterisk indicate stellar cluster objects listed in \cite{DI}, a plus sign shows objects that are listed in \cite{bica} and the circle indicates those which are listed in \cite{bd}. The star locates FAUST 392. Each of the objects are indexed in Table~\ref{tab:obs_dem171}. Contours are 2$^x$\% of peak intensity ($x$=2-6). Intensity units shown on the colour wedge are in Rayleighs.\label{fig:dem171}}
\end{figure}

The feature shown in Figure~\ref{fig:dem171} is most easily interpreted as an expanding and internally-heated stellar wind or SN shell. 

DEM171 is known from \cite{meaburn} and others \citep{graham,parker_aao}, who made some speculation regarding the ionisation source. There are a number of candidate heating sources which are internal to this shell \citep{bica} and candidates include FAUST 392 (shown with the black star) which may be an 08-type or WR star. A comparison of the \ha\ from DEM171 with an easily identifiable expanding \hi\ shell has already been made by \cite{mul_hi}.

\begin{table*}
\begin{center}
\caption{\ha\ objects and candidate heating mechanisms. Table~\ref{tab:faust} contains a list of FAUST objects that are associated with observed \ha\ nebulosity in the Magellanic Bridge.}\label{tab:ionisation_sources}
\begin{tabular}{lllll}\hline
Object & Angular Size                & Description &RA, Dec &Cand. energy source\\
       & [arcmin] $\pm$6 arcsec      &              &(J2000)&               \\\hline\hline
DEM 171$^{*}$&8.4$\times$7.8&Complete \ha\ ring &02\hr\ 07\m\ 51.7\s $-$74\dd\ 44$'$ 08.0$"$ &FAUST 392\\
                        &&&                                              &ICA 13 Assoc.\\
                  	&&&                                              &BS95 DI\\
MP1$-$N$^{*}$&0.38 (FWHM)&North patch in group&01\hr\ 56\m\ 34.4\s $-$74\dd\ 17$'$ 00.9$"$ &BS95 215\\
MP1$-$S$^{*}$&0.87 (FWHM)&South patch in group&01\hr\ 56\m\ 45.3\s $-$74\dd\ 13$'$ 10.6$"$ &WG 8\\
MP1$-$E$^{*}$&0.95 (FWHM)&East patch in group&01\hr\ 57\m\ 01.4\s $-$74\dd\ 15$'$ 43.7$"$ &BS95 216\\
                  	&&&                                            			   &BS95 217\\
MP1$-$W$^{*}$&0.80 (FWHM)&West patch in group&01\hr\ 56\m\ 29.6\s $-$74\dd\ 14$'$ 59.4$"$ &FAUST 339\\
MP2$^{\dag}$&2.7 (E-W)&E-W filament &01\hr\ 56\m\ 37.3\s $-$74\dd\ 30$'$ 55.0$"$ & N/A\\
MP3$^{*}$&16.1$\times$18.0&Extended Patch &01\hr\ 50\m\ 54.8\s $-$74\dd\ 10$'$ 54.8$"$ &FAUST 318\\
                        &&&                                              &WG 5\\
MP4n$^{*}$&7.0$\times$8.6&Crescent &01\hr\ 49\m\ 42.1\s $-$74\dd\ 36$'$ 55.3$"$ &FAUST 313\\
                        &&&                                              &WG 2\\
MP4s$^{*}$&0.75 (FWHM)&Rounded patch &01\hr\ 49\m\ 28.4\s $-$74\dd\ 39$'$ 09.2$"$ &BS95 200\\
MP5n$^{*}$&1.7$\times$1.0&North patch of two &01\hr\ 43\m\ 26.6\s $-$74\dd\ 31$'$ 43.4$"$ &\\
MP5s$^{*}$&1.1$\times$1.6&South patch of two&01\hr\ 43\m\ 57.5\s $-$74\dd\ 33$'$ 37.9$"$ &\\
MP6n$^{*}$&7.7 (length of HP)&Filament in complex&01\hr\ 40\m\ 54.7\s $-$73\dd\ 51$'$ 28.8$"$ &FAUST 279\\
MP6s$^{*}$&4.8 (length of HP)&Filament in complex&01\hr\ 41\m\ 01.8\s $-$73\dd\ 54$'$ 22.4$"$ &FAUST 279\\
\hline
\end{tabular}
\medskip\\
\end{center}
$^{*}$Radiatively Heated
$^{\dag}$Collisionally heated (?)\\
\end{table*}

\begin{table}
\begin{center}
\caption{Candidate ionisation sources within DEM171}\label{tab:obs_dem171}
\begin{tabular}{ll}\hline
Index&Object\\\hline\hline
1&BS95 230\\
2&DI91 437\\
3&BS95 228\\
4&DI91 453\\
5&DI91 427\\
6&DI91 380\\
7&FAUST 392\\
8&DI91 474\\
9&DI91 363\\
10&DI91 374\\
11&DI91 337\\
12&ICA 13\\
13&DI91 376\\
14&DI91 384\\
15&DI91 42\\
16&DI91 338\\
17&DI91 334\\
18&DI91 353\\
19&DI91 523\\
20&DI91 523\\
\hline
\end{tabular}
\end{center}
\end{table}

\begin{table}
\begin{center}
\caption{FAUST objects that are associated with observed \ha\ nebulosity}\label{tab:faust}
\begin{tabular}{cc}\hline
Associated&FAUST \\
\ha\ object&object number\\\hline\hline
MP1W&392\\
MP3&318\\
MP4&313\\
MP6&279,265,265\\
\hline
\end{tabular}
\end{center}
\end{table}

\subsection{Cluster of 4 small, diffuse features: MP1N,S,E,W}
\begin{figure}
  \includegraphics[scale=.3, angle=-90]{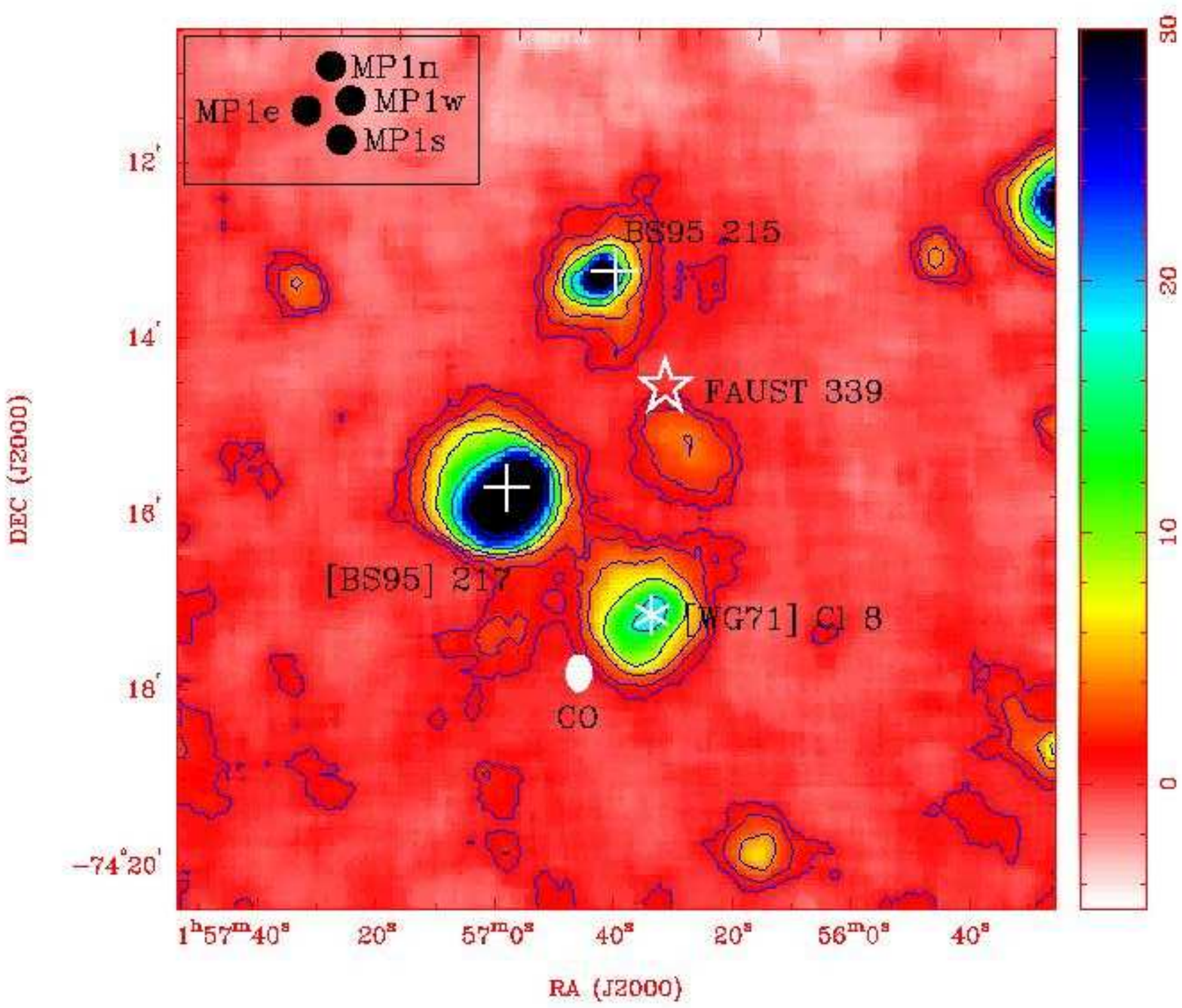}
 \caption{UKST H-alpha data, showing the positions of four, perhaps associated \ha\ emission features; MP1N,S,E,W. This region is spatially correlated with $^{12}$CO(1-0)  line emission \citep{mul_co}, which is shown with the white ellipse (representative of the ~35$"$ beam of the Mopra mm telescope). The relative positions of potential heating sources; FAUST 339 (star), BS 215, 217 (plus symbol) and WG 8 (asterisk symbol) are also shown. The legend in the top left indicates the adopted naming convention. Contours are 2$^x$\% of peak intensity ($x$=0-6). Intensity nits shown on the colour wedge are in Rayleighs.}\label{fig:4blob}
\end{figure}

In the absence of evidence to the contrary, we will assume that the unlikely spatial coincidence of the the four regions shown in Figure~\ref{fig:4blob} suggests that they are associated. They may have formed from a common gas cloud, or that they represent sites of secondary starformation (perhaps) catalysed through the evolution of a stellar wind or SN shell. The four objects are listed in Table 1 as MP1N,S,E,W. See \cite{mul_co} for a comparison and description of the \hi\ and dust emission associated with these objects. With the exception of MP1W, each of the \ha\ regions have plausible internal heating mechanisms which are elaborated below.

\subsubsection{MP1N}
MP1N is clearly associated with the BS95 cluster 215 \citep{bica}. This is very compact cluster and is clearly a strong source of high energy radiation. There appears to be little dust and other \hi\ associated with this object, implying that it is a more evolved, although still rather young group. It has a less diffuse morphology in \ha\ than does MP1S, implying that some of the nursery gas has been dispersed. 

\subsubsection{MP1S}
MP1S also has a clear association with a stellar group WG71 8 \citep{wg71}. The \ha\ morphology is a little more extended than the northern feature (MP1N), although it also appears to be a more compact and less well populated group. This object is most closely related to the site of the first detection of $^{12}$CO(1-0) in emission within the Bridge \citep{mul_co}. The extent of the CO at this position has been explored by \cite{mul_co} and has been shown to be very compact \citep[R$<$16 pc][]{mul_co}. Weak CO has also been found at other locations in the Magellanic Bridge \citep{mizuno}. The CO detected at MP1S appears to reside in a local overabundance of \hi\ which would shield the CO from any otherwise destructive radiation which is emitted by WG71 8. This process is evidenced by the excited \ha\ emission shown here.

\subsubsection{MP1E}
MP1E is spatially coincident with OB associations BS95 217 and 216, although a reference to \cite{bica} shows that BS95 216 is a sub-cluster within the larger BS95 217. This object is also listed in the IRAS catalogue (IRAS 01562-7430). The strong radiation at Infrared wavelengths implies a significant warm dust component. 

\subsubsection{MP1W}
This object is not coincident with any ionising sources that are visible in the optical, but is most closely geometrically related to the UV source FAUST 339. The somewhat axi-symmetric morphology of MP1W is not particularly consistent with heating via an external source and as such, this region may harbour an as-yet invisible site of starformation.

\subsection{East-West Filament: MP2}
\begin{figure*}
  \centerline{\includegraphics[scale=.3, angle=-90]{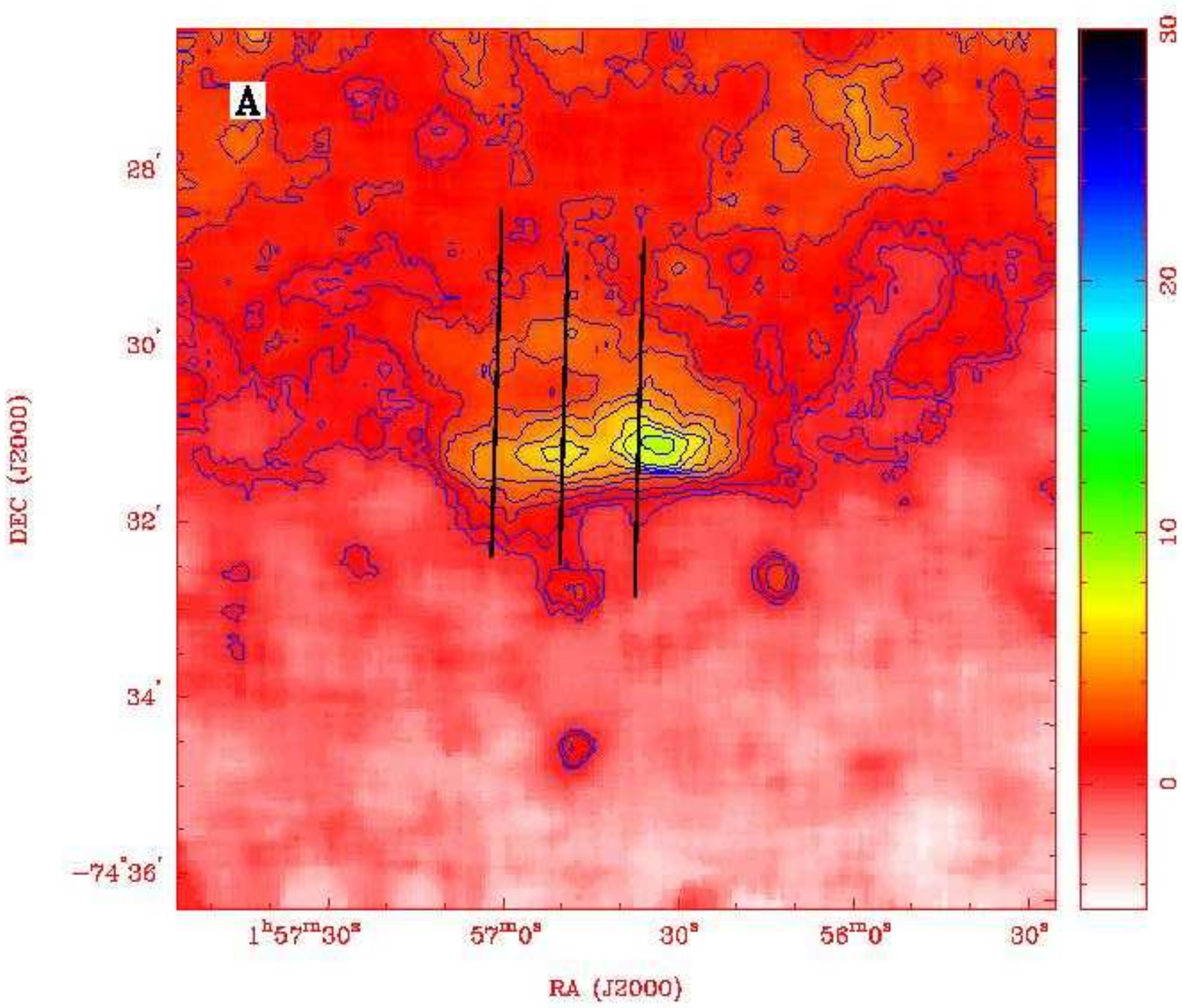}}
 \centerline{
 \includegraphics[scale=.15, angle=0]{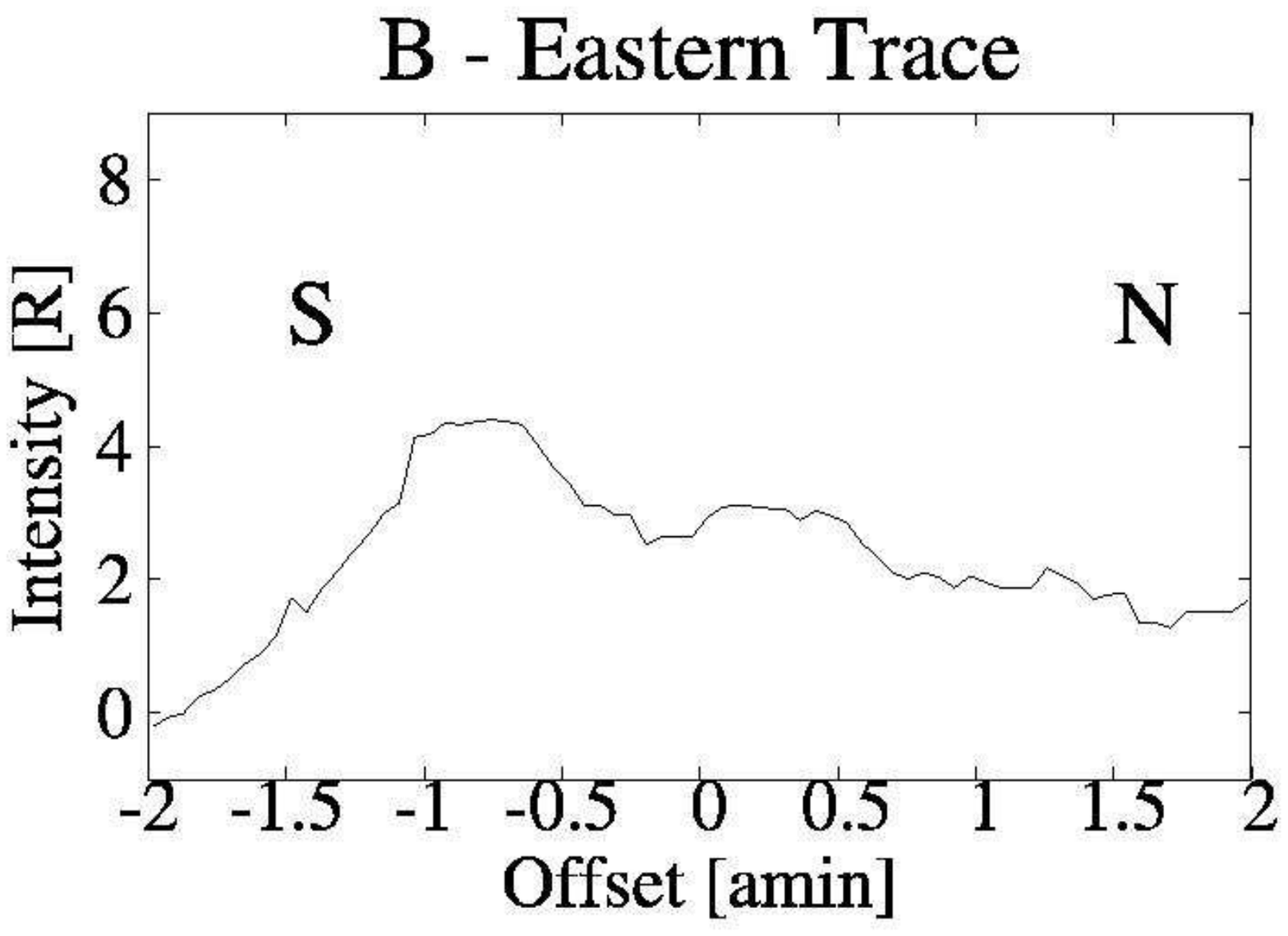}
 \includegraphics[scale=.15, angle=0]{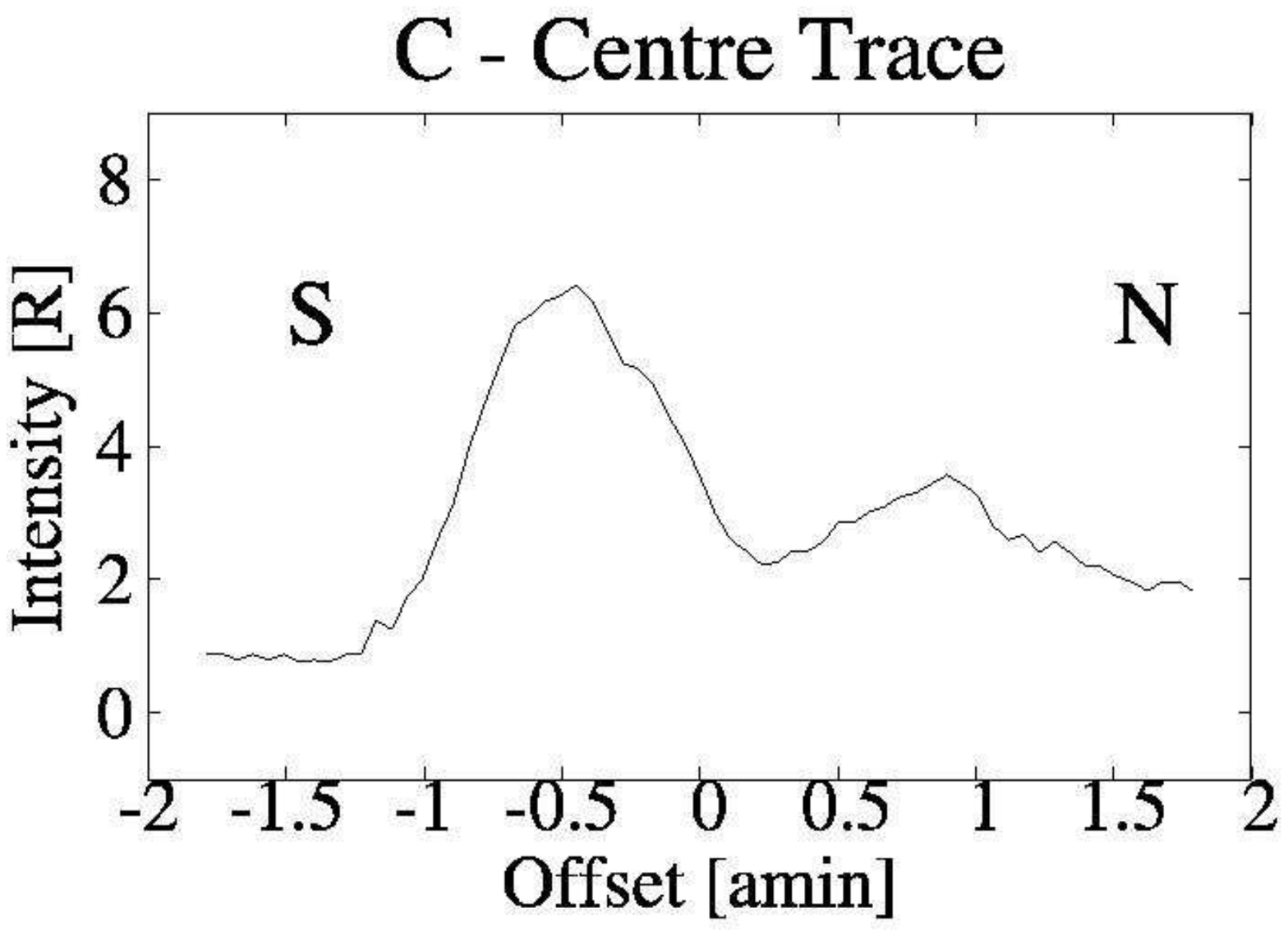}
 \includegraphics[scale=.15, angle=0]{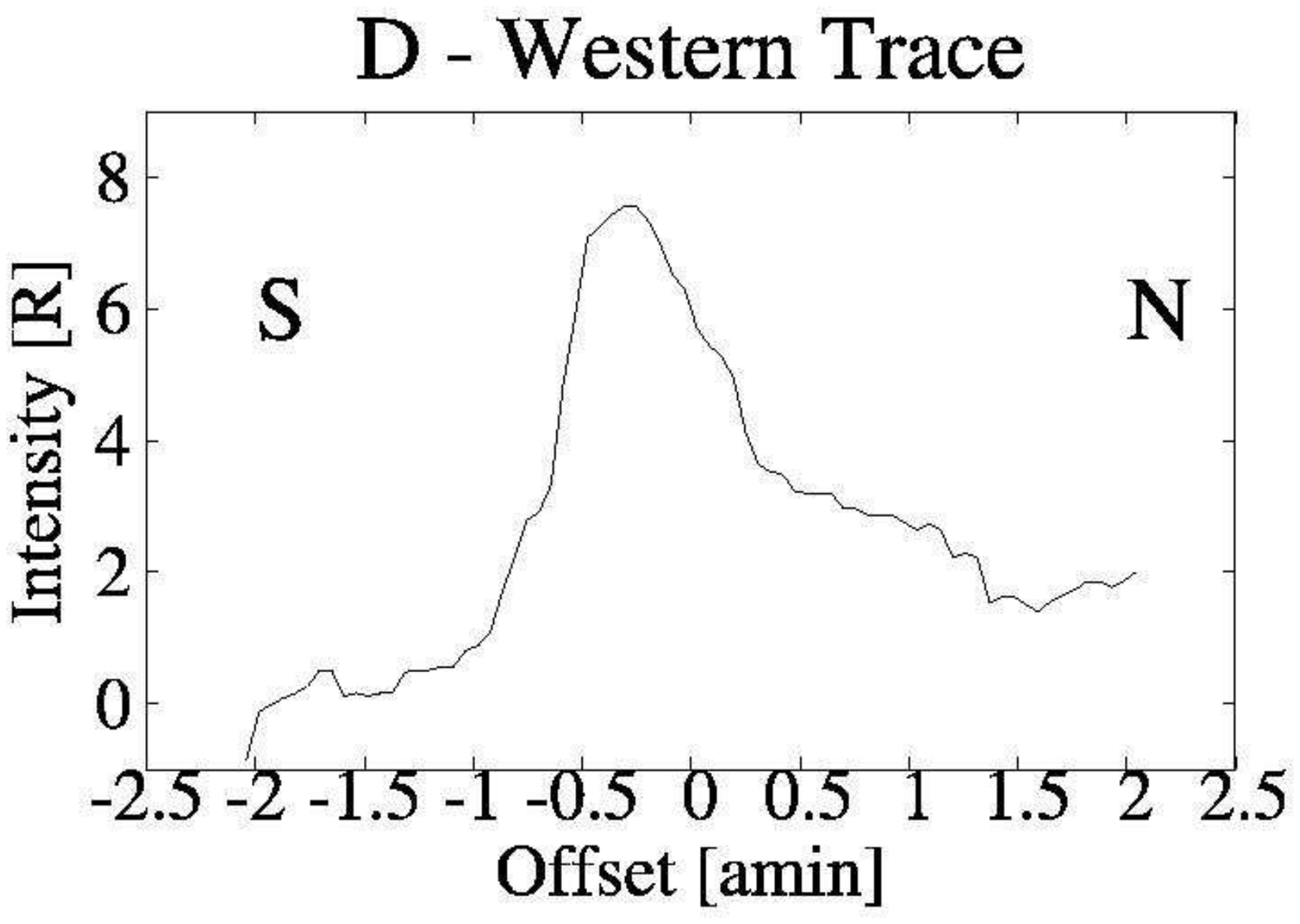}}

  \caption{(\emph{Panel A}) UKST H-alpha data, showing an E-W filament (MP2). This is the only detected \ha\ feature in the Bridge that does not have a readily detectable ionisation mechanism.
  Contours are 0.5,1:10+1, 12:20+2 R. Intensity units shown on the colour wedge are in Rayleighs.\newline
  (\emph{Panels B-D}) Three profiles, shown as the tracers in the top panel. These profiles show a steeply rising increase in emission from the south, which is suggestive of a shock front.}
  \label{fig:e_w_filament}
\end{figure*}

Figure~\ref{fig:e_w_filament} (\emph{Panel A}) shows a relatively unusual emission object: a region of rather bright \ha\ emission, without having any detectable candidate heating source. This feature has a shock-front morphology, in that the intensity contours are tightly clustered on one side and much more separate on the opposite side. Such a structure is suggestive of compression of the ISM by a shock, formed perhaps by sheer/collisional interaction between Bridge-Bridge ISM, or between the Bridge and external material. Overlaid on Figure~\ref{fig:e_w_filament}A are three vertical lines, which indicate the traces shown in Figure~\ref{fig:e_w_filament}B-D. The steeply-rising profile from the south is clear in these plots. High sensitivity spectral information are necessary to test for molecular and ionic shock tracers to confirm the shock-front scenario.

\subsection{Large diffuse Patch: MP3}
\begin{figure}
\begin{center}
  \includegraphics[scale=.3, angle=-90]{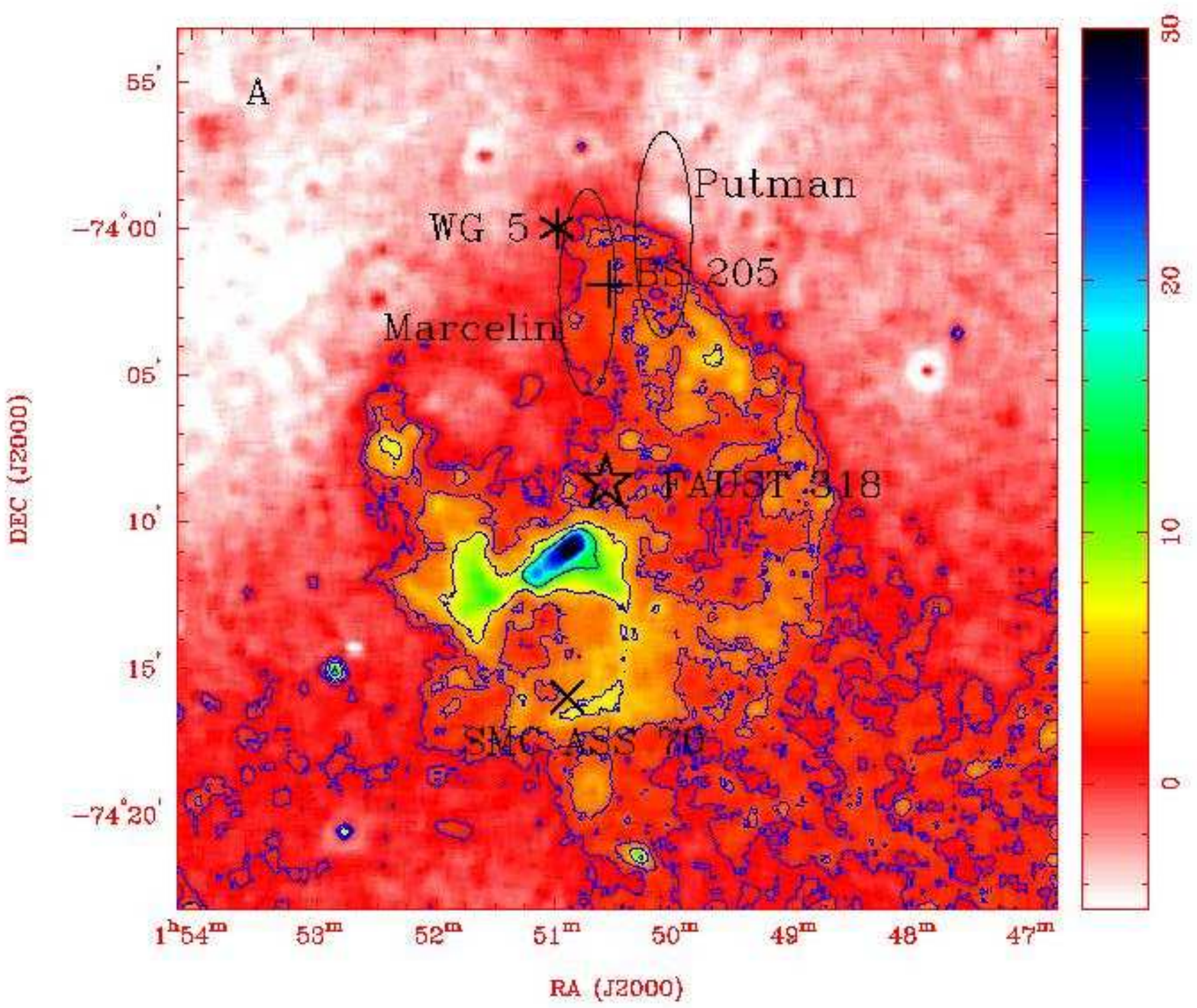}
  \includegraphics[scale=.3, angle=-90]{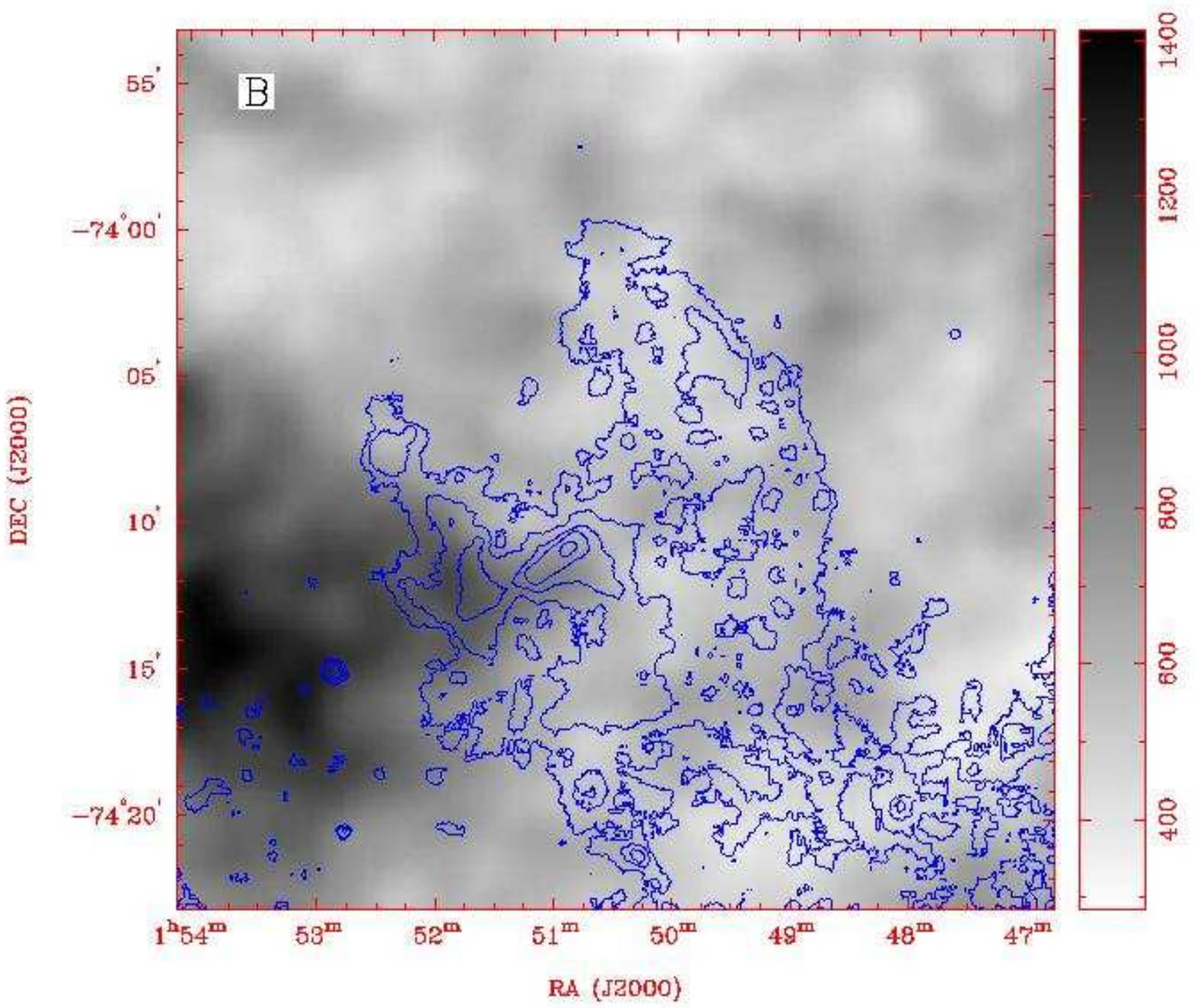}
 \end{center}
  \caption{(\emph{Panel A}) UKST H-alpha data, showing the diffuse patch (MP3).  Ellipses show positions of \ha\ observations by \cite{mbg} and \cite{putman_ha}. The star locates FAUST 318, the plus sign shows the position of the OB cluster BS 205 \citep{bica}. The cross shows the position of the stellar cluster WG 5 \citep{wg71} and the asterisk shows the position of the SMC stellar association 70 \citep{hodge}. Contours are 2$^x$\% of peak intensity ($x$=2-6). Intensity units shown on the colour wedge are in Rayleighs.\newline
(\emph{Panel B}) Greyscale: Integrated \hi\ line emission of region associated with the \ha\ Diffuse Patch. The \ha\ region is shown as contours. Brightness units shown on the colour wedge are in Kelvin }\label{fig:patch}
\end{figure}

This approximately circular \ha\ patch is shown in Figure~\ref{fig:patch}A. It subtends $\sim$20$'$ at the fullest extent and it is the largest \ha\ region yet identified in the Bridge. Despite the large size, it is difficult to convincingly associate the \ha\ morphology with the brightness distribution of \hi. Figure~\ref{fig:patch}B shows the \ha\ overlaid as contours on \hi\ and apart from a weak geometric correlation of an \hi\ filament which appears to follow the centre of the \ha\ region, there are no strong morphological reasons to suggest that the \hi\ line emission and the \ha\ are closely related.

This feature has been noted in very early work by \cite{jmo} who identified it as the brightest region of \ha\ in the Bridge. These authors measured a brightness for this region of 2.5x10$^{-9}$JM$^{-2}$s$^{-1}$sr$^{-1}$ ($\sim$10 R).  After smoothing to the resolutions of \cite{jmo}, we measure a peak brightness of $\sim$5 R, although we note that this feature falls exactly upon a low-level film artefact which would significantly affect the baseline-subtraction process.

Even later followup observations, by \cite{mbg} and then \cite{putman_ha} failed to point at the position of peak \ha\ brightness for this feature, although they were using UV as a trace for energetic processes \citep{courtes}. The positions and areas of observations by \cite{mbg,putman_ha} are shown as ellipses on Figure~\ref{fig:patch}. These authors report an emission measure of 4 and 3.2 R (respectively) and the 400\% error here in our approximate measured values highlights the difficulties in extracting low-level and wide-field emission from the UKST dataset in the absence of strong extended reference observations.

Central to this feature is the object FAUST 318. At this early stage, based on the morphology of the \ha\ feature and without a means to fully probe the ionisation flux of the \ha\ region, we suggest that FAUST 318 is the source responsible for generating the observed \ha\ emission. 

\subsection{Crescent: MP4} 
This feature is one of two identifiable \ha\ ring-like features in the Bridge. The object FAUST 313 is located nearby the brightest \ha\ emission from the crescent and is therefore the mostly likely heating source which gives rise to the observed emission. There are no identifiable, centrally-located candidate heating sources. An additional, compact \ha\ emission region (MP4a) is detected to the south west of the crescent. This smaller region is exactly coincident with an OB association BS95 200 \citep{bica}. Of interest is the coincidence also with IRAS 01489-7452 and the blue cluster WG 3. These may represent different detections of the same object, given the pointing errors of the FAUST and IRAS instruments.

\subsection{Two small, diffuse features: MP5}
\begin{figure}
\begin{center}
  \includegraphics[scale=.3, angle=-90]{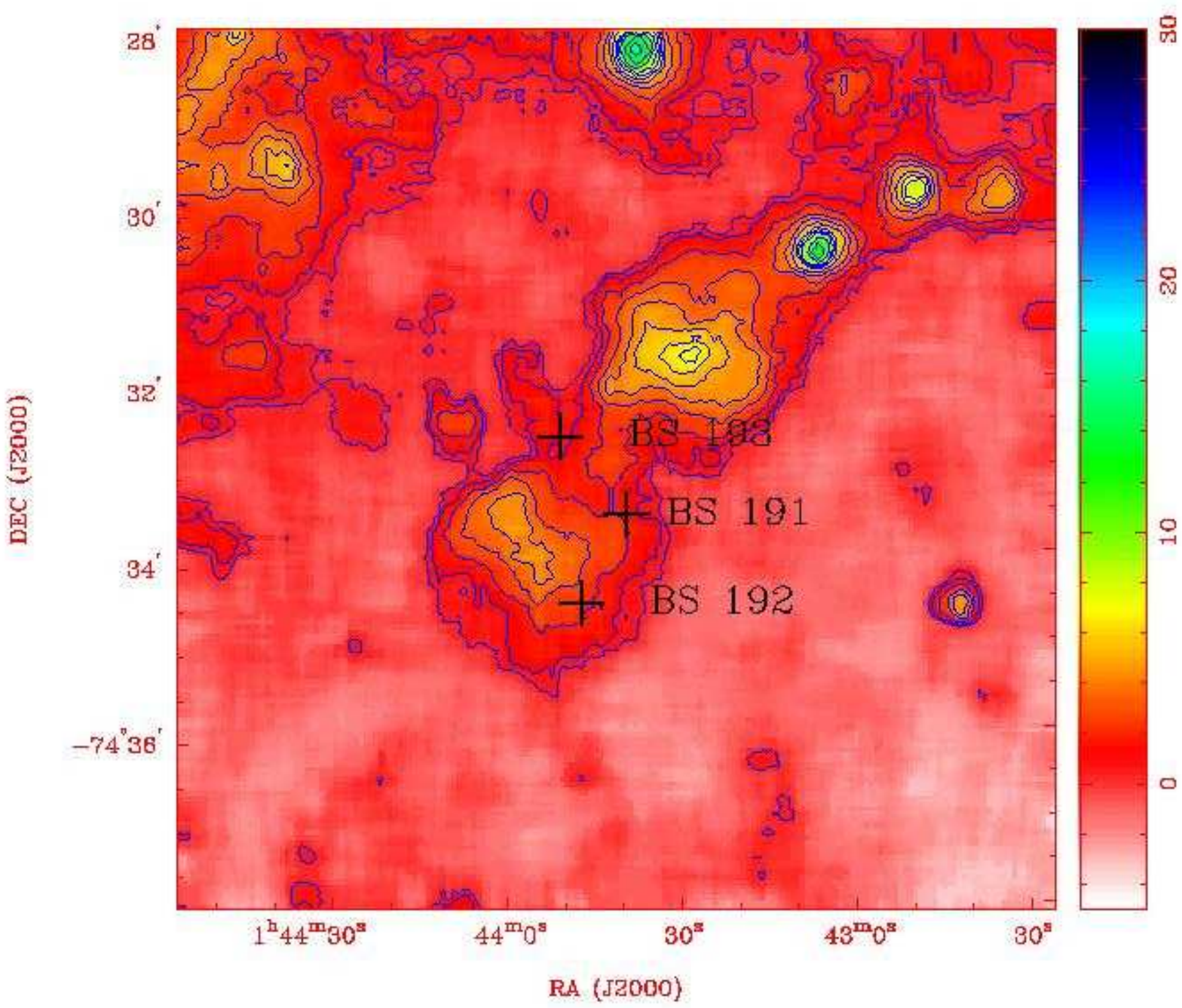}
\end{center}
  \caption{UKST H-alpha data, showing two diffuse circular features (MP5). OB association catalogue objects are identified with a black plus sign. There is a general, but not very close correlation of the OB associations with the peaks of the \ha\ emission. Contours are 2$^x$\% of peak intensity ($x$=0-6). Intensity units shown on the colour wedge are in Rayleighs.}\label{fig:2blobs}
\end{figure}

The two circular regions of \ha\ emission (MP5n, MP5s) are shown in Figure~\ref{fig:2blobs}. The adjacent regions are relatively faint, being only a few Rayleighs in total emission. These objects are within a few tens of pc (projected line of sight) to optically identified OB associations \citep{bica}, although the exact alignment of the associations does not correspond to the centre of the emission. The geometry of the \ha\ and OB associations imply that the OB associations may not be responsible for the observed ionisation recombination, but may represent members of a common starforming region. A candidate scenario for these emission regions is that they are being ionised by embedded and obscured star formation, however, these sites are not apparent in 60$\mu$m dust-emission maps which would help to support the internal radiative heating scenario.

\subsection{Filament complex: MP6}
\begin{figure}
  \includegraphics[scale=.3, angle=-90]{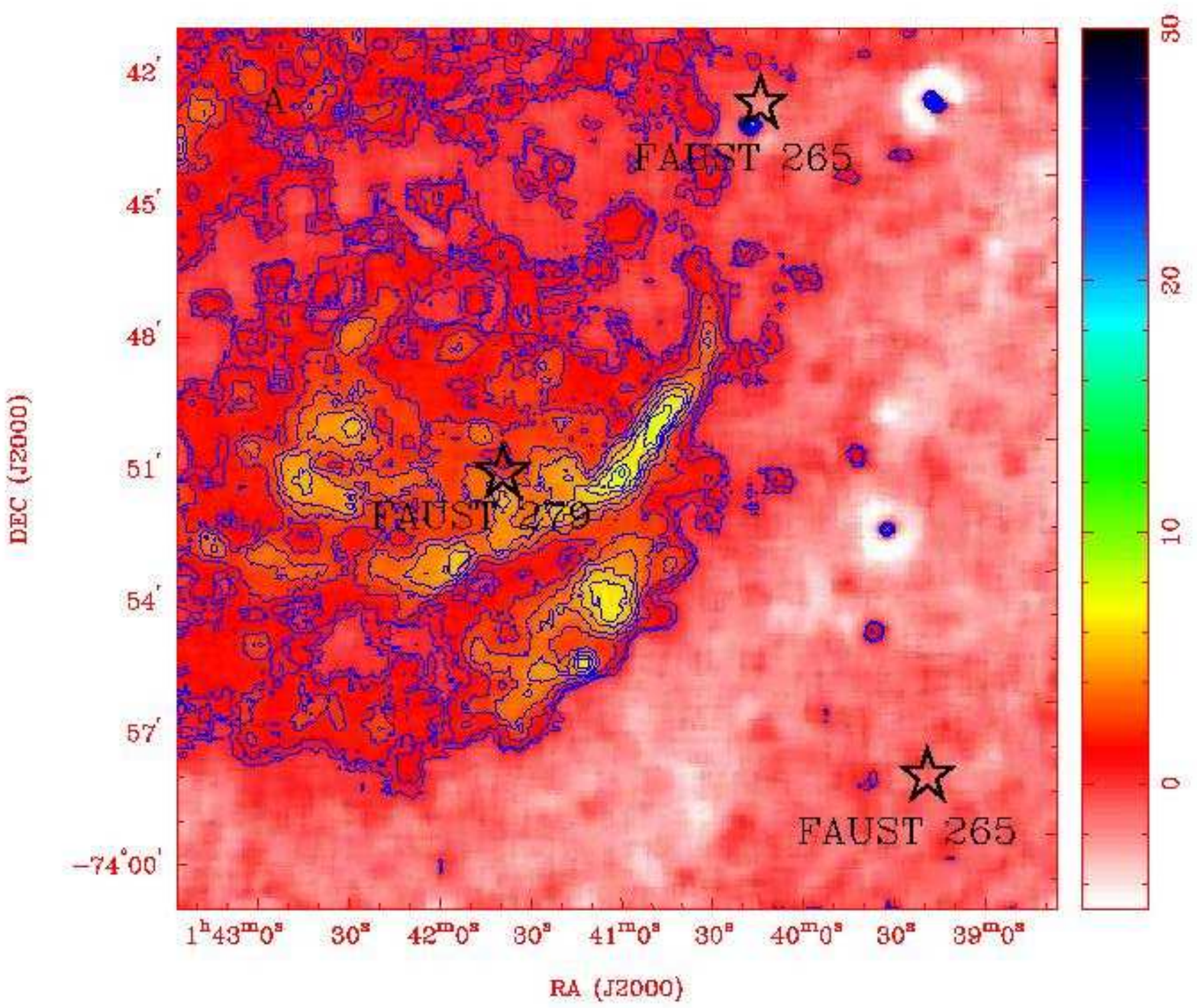}
  \includegraphics[scale=.3, angle=-90]{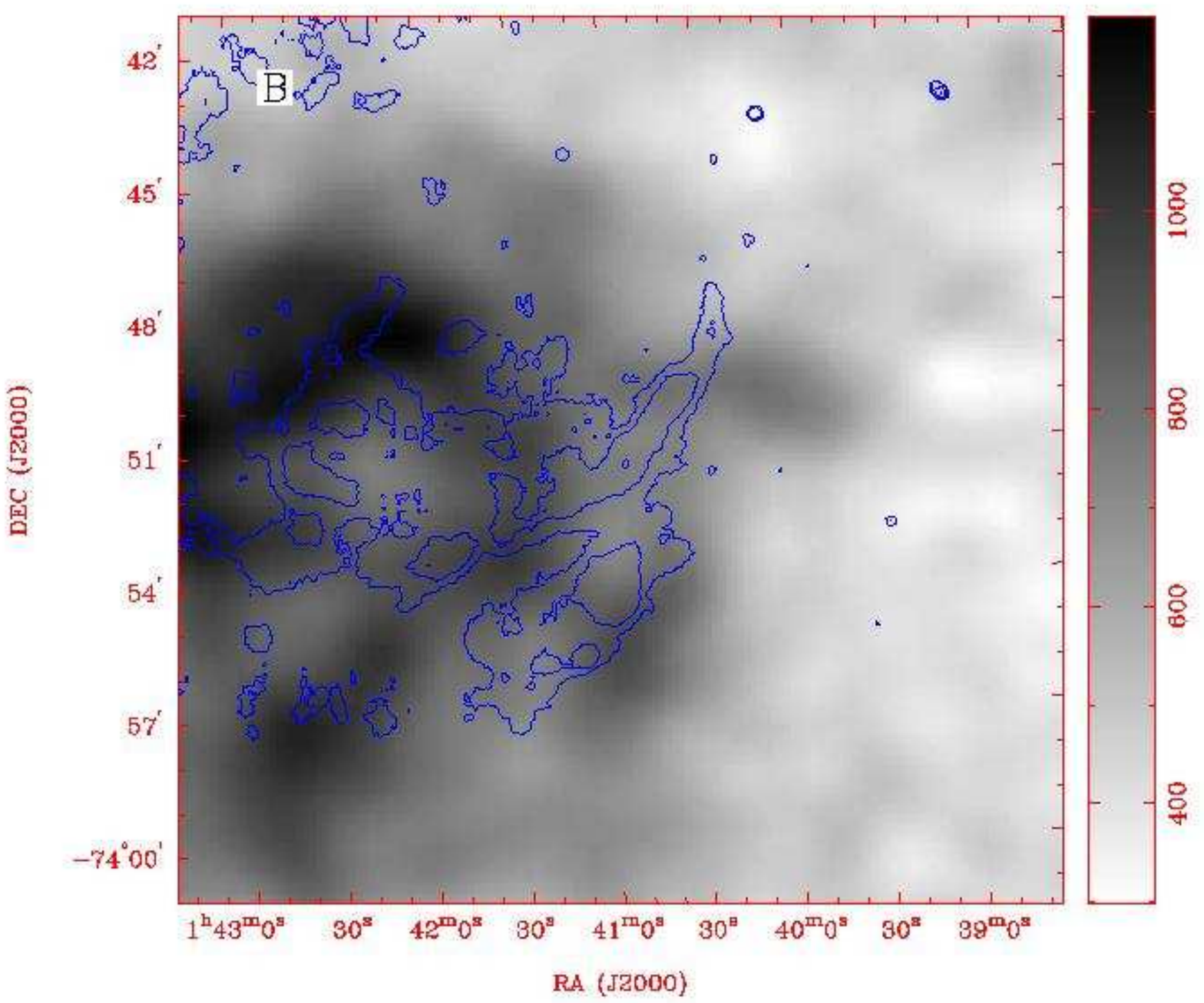}
  \caption{(\emph{Panel A}) UKST H-alpha data, showing a filamentary complex (MP6).  The positions of three FAUST objects are shown with stars. Contours are 2$^x$\% of peak intensity ($x$=2-6). Intensity units shown on the colour wedge are in Rayleighs.\newline 
(\emph{Panel B}) Greyscale: Integrated \hi\ line emission of region associated with the filaments. \ha\ emission is overlaid as contours with intervals of 0.5,1:10+1, 12:20+2 R. The Greyscale brightness shown on the colour wedge are in Kelvin.
}\label{fig:f_filaments}
\end{figure}

These \ha\ filaments shown in Figure~\ref{fig:f_filaments} only loosely correlate with the large-scale contours of the \hi\ brightness. The more diffuse \ha\ is associated with large integrated \hi\ brightness, however, the brightest edges of the \ha\ filaments are uncorrelated with the \hi. Again, the proximity of these features to a UV source suggest radiative heating of the ISM. FAUST 279 is the only object which is located centrally to the \ha\ emission.

\section{Scale-Energy Estimates:}
The observed \ha\ emission is a direct signpost of the quantity of energy that is capable of interfering with the turbulent structure of the ISM. We can test the assumption that energetic processes arising from the formation and death of stars will shape the Bridge ISM, by estimating power from the observed \ha\ nebulosities. 
We observe that the brightest parts of the \ha\ regions (i.e. with power $>$ 10 R) subtend a few arcmin ($\sim$10 pc). Using R = 2.71$\times$10$^{-7}$ erg s$^{-1}$cm$^{-2}$sr$^{-1}$, we can sum the total power from each of the \ha\ features at this scale to yield: \\
$\sum P$(\ha) = $\sim$10$^{42}$ erg (T = 5 Myr;  D $\sim$60 kpc). \\

This order-of-magnitude estimate is substantially lower than the typical energy output of any early-type massive star. Under the assumption that the observed \ha\ emission is directly related to energetic stellar processes that are capable of perturbing the turbulent ISM at these pc scales, we can conclude that in the case of the Magellanic Bridge, the stellar population does not contribute significantly to the turbulent energy balance of the ISM. The lack of emission which is not easily attributable to compact candidate heating sources also suggests that the turbulent ISM of the Bridge is not significantly perturbed by other high-power (for example; collisional) processes.

\section{Conclusions}
The \ha\ observations presented here reveal a number of weak \ha\ features located in the tidal Magellanic Bridge. The low emission measures of all the features suggest that the ISM within the Bridge is not being substantially mixed by the deposition of energy via turbulent motions, stellar winds and SNe. The lack of substantially energetic processes in turn imply that the ISM in the Bridge is largely a purely turbulent feature, from the largest scales of a few kpc, down to the scales probed by this report; of a few pc. We conclude that the Magellanic Bridge is therefore the largest and closest tidal feature that harbours a pure turbulent structure that is available for study.

With the exception of one feature, all of the observed emission regions appear to be close in projection to bright sources of UV radiation. We have been restricted in the accuracy of this analysis by an insecure calibration scheme. Obtaining well calibrated \ha\ spectral line data will be important in providing both power and kinematic information and we will then be in a more tenable position to speculate reliably on the mechanisms giving rise to the observed \ha\ features. In addition, measurements of ionic species, such as [S${\sc II}$] which are indicators of shock-heating, will provide a means to clearly discriminate between viable ionisation schemes.  We will develop this project further and address these questions.

%



\end{document}